\documentstyle[preprint,aps,epsf]{revtex}

\tighten
\begin{document}
\bibliographystyle{/usr/share/texmf/tex/latex/revtex/prsty}
\draft
\title{ $^9$Be: A gateway nucleus into heavier nuclei?  }

\author{ Y. Koike}

\address{Science Research Center, Hosei University, Chiyoda, Tokyo
102, Japan  }

\date{\today}

\maketitle

\begin{abstract}

Level structure of $^9$Be is investigated  calculating  poles of 
the $\alpha$$\alpha$n three-body 
scattering amplitude in the complex energy plane.
  We find 10 levels,
which make rotational bands in both positive and negative parities.
All of them, except 7/2$^+$, have been observed experimentally and the 
corresponding positions and widths reasonably well matched.  
The 7/2$^+$ has a large width, and
therefore has not been found, but may cause a noise in the experiment.  
No other levels are needed to explain the existing 
experimental observables below 11.28 MeV, where the 9th excited state exists. 
We definitely assign it 9/2$^-$.    
\end{abstract}

\narrowtext

$^9$Be is a mysterious nucleus.  
The number of levels of the two Be isotopes, $^8$Be and $^9$Be, 
reported in the standard compilation[SC]\cite{comp},
is considerably larger compared to lighter nuclei(A$\le$7).
The number of energy levels in $^9$Be  
counts 10 already below 12 MeV, whereas only 3 levels are found in $^8$Be in 
the same energy region. While 3 levels of $^8$Be are simply rotational 
levels($\ell$=0,2,4), the nuclear structure of these 10 levels is not
trivial. 
Thus $^9$Be seems a gateway into heavier nuclei where levels are rich and 
non-trivial of low excitation.

All levels except for the ground state are resonances. They usually have 
decay widths considerably large, which makes the experiment difficult. 
Among these ten levels, only the 6 lowest lying levels below 5 MeV are
experimentally understood without any ambiguity.  
Although the level ordering is complicated, these 6 levels are 
low spin states 1/2$^\pm$-5/2$^\pm$.  
Two different categories of the nuclear structure are possible 
in the energy region of 5-12 MeV according to the shell model prediction.
One of them is a second 
level of low spin states, such as 3/2$^-_2$, and the other is high 
spin states such as 7/2$^\pm$.   They make the level structure of this nucleus 
ambiguous.
Two correlated new measurements[NM]
\cite{ex1,ex2}  strongly contradict SC in this region.

According to NM only one definite bump can be seen at 
6.7 MeV where SC 
reports a 7/2$^-$ level, whereas NM
report two levels: 7/2$^-$ at 6.38, and 9/2$^+$ at 6.76 MeV.  They 
find it necessary to introduce two levels in order to explain consistently
the angular distribution of the bump both in the cross section and in the 
analyzing power .  
SC reports a level 1/2$^-_2$ at 7.94
MeV, which is not reported in NM.   
Instead, NM
report a new level 3/2$^-_2$ at 5.59 MeV  which is not reported
in SC.   These are the second levels of the same spin parity state 
reported in the experiment.  Thus the evidence of second levels is 
rather weak.  They are located on both sides of the 
6.7 MeV bump and have widths of $\sim 1$ MeV.  

The first clear peak above this region 
is at 11.28 MeV, where no spin-parity assignments are given in 
SC whereas a weak assignment of 7/2$^+$ is given 
in NM.

As a result, a new compilation in a preliminary version\cite{compn} lists 
some low lying
levels of $^9$Be with very few experimental evidences.  Spin parity assignments,
in considerable cases, are unclear.  

Many theoretical works\cite{theo1}-\cite{enyo} have been done 
on this nucleus, but the situation is still
rather confusing for lack of experimental knowledge.
The most successful theoretical work, so far, was done by 
Arai {\it et al.}\cite{theo1}, who
worked in the framework of the microscopic three-body model and used the complex
scaling method to predict not only levels but also the widths.  
The theoretical prediction by Arai {\it et al.} explained all low lying levels
,with their positions and widths, 
except for the first excited level 1/2$^+$.  However, they did not resolve the
confusion described above.  

All excited states below 12 MeV decays only into $\alpha$$\alpha$n three 
particles. 
There are no two particle decaying channels.   This means levels should be 
described in close relation to three-particle scattering.  
Faddeev theory gives a rigorous tool to describe three-body dynamics.  It 
is well established that the low energy six-nucleon 
system($^6$He and $^6$Li) can be described very well by Faddeev theory below the $\alpha$ particle breakup threshold within a framework 
of an $\alpha$NN three-body model\cite{leh}-\cite{esk}.  
The deuteron breakup reaction on the $\alpha$ particle is widely 
explained\cite{koi1}, for which no other approach has been successful. 
Resonance poles in the complex energy plane were also investigated\cite{esk}.

Here we adopt a similar approach introducing a three-body 
$\alpha$$\alpha$n model. We look for levels as poles in the three-body
scattering amplitude.
The approach is in many ways phenomenological,
but proper treatment of three-body
dynamics gives a solid base to describe all possible low lying
resonance states that couple strongly to the $\alpha\alpha$n decaying channel. 
The thresholds corresponding to various channels 
in the scattering amplitude are given explicitly by the two-body
interactions adopted. 
In order to
make sure that the existence of poles of the scattering amplitude is 
potental independent, we use different two-body interactions 
which have been reported in the past.  They are parametrized such
that they reproduce phase shifts in the $\alpha$$\alpha$ and $\alpha$n 
scattering reasonably well, and therefore describe all thresholds correctly. 

We use homogenous AGS equations\cite{ags} and calculate the 
eigenvalue $\lambda(E)$ with a complex energie $E$:  
\begin{equation}
  \chi_c(q,E)= \lambda(E) \sum_{c'} \int_0^\infty  q'^2 dq Z_{cc'} (q,q':E) \tau_{c'}
     (E-{q'^2 \over 2\mu} )  \chi_{c'}(q',E).  
\end{equation}
 If $\lambda(E) = 1$ at a certain 
energy $E$, there is a pole in the scattering amplitude. 
Since the poles are in the complex energy plane, a special treatment is needed. 
We adopt the contour deformation method which is well known as a technique to
solve the scattering equations.  The method can be found in ref.\cite{esk}.
Instead of taking a straight line for $q$, we adopt a straight contour up to a
certain point, then run parallel to the real axis.  This is necessary because
the form factor used in the present work is more sophisticated than in 
ref.\cite{esk}.

There are several thresholds in the scattering amplitude.  
In addition to the breakup threshold, there are $\alpha$$\alpha$ and $\alpha$n
thresholds from their respective interactions in $\ell$=0,2,4 and p$_{3/2}$, 
p$_{1/2}$.
These thresholds cause singularities in the energy plane. 
Since levels in two-body subsystems are resonances, the 
singularities, where the cuts start toward the infinity, 
are complex.
In the present calculation, we choose a contour which makes all the cuts start
from corresponding singularities downward parallel to the imaginary axis,
before running toward positive infinity parallel to the real axis.  

In order to use this technique which requires 
one dimensional integral equations, we use separable potentials in two-body
subsystems.   If the potential is local, we make a separable expansion.
Two-body interactions we use are as follows.  To describe the more important
$\alpha$$\alpha$ interaction, we use standard potentials reported in the 
literature.  We use Ali-Bodmer(AB) potential\cite{ab} as well as 
Chien-Brown(CB) potential\cite{cb}.
A separable expansion is done for both potentials and convergence check has been
done carefully both in two and three-body systems.
We also use a separable UIM potential derived from RGM theory\cite{uim}.  
For the $\alpha$n interaction, we adopt simple Yamaguchi type separable 
potentials.  We use a repulsive potential for s-wave, while attractive 
potentials for p$_{3/2}$ and p$_{1/2}$ waves.  
With potentials described above, we can try different combinations. 
Two of them, with AB and CB, have been successfully used by Cravo\cite{cravo} 
in the ground state calculation $^9$Be, treating Coulomb interaction properly.

In order to keep the calculation possible above scattering thresholds, we
neglect the Coulomb interaction between the $\alpha$'s, which  causes a 
shift in the energy levels as well as minor change in the width.
The shift is rather small, about 1.8 MeV, and is expected to remain 
unchanged\cite{pen}  for all the levels, keeping the level spacing the same.

The binding energy of $^9$Be calculated with UIM is 
-3.32 MeV.  Taking 1.8 MeV as the energy shift due to
the Coulomb interaction\cite{cravo}, one gets a good fit to
the experiment.   
Therefore, throughout this report, we show our result with UIM.  
Assuming the Coulomb energy shift constant, we give excitation energies
from the calculated ground state energy. 
We should notice that we find all potentials described above give more 
or less the same results for all levels given below; general features 
described here are potential independent.  

Table 1 shows our results and the results  by Arai {\it et al.} for 6 well 
understood levels together with the experiment.  Both calculations 
give a good fit to the experiment except for the first excited state 1/2$^+$.  
We could not obtain the first excited state(1/2$^+$) pole with the contour used 
in the present calculation.  However, the
eigenvalue in the hemogenous equations shows that there is a pole in the 
complex energy plane.  Therefore, 
it should be a virtual state pole, as pointed out by Efros and Bang\cite{efros},
which we are now looking for. 
 
In the absence of an $\alpha$n $\ell$$\cdot$s force, 
two levels, in general, are degenerate.  They are, for instance, 
 3/2$^+$ and 5/2$^+$ in the
positive parity, or 3/2$^-$ and 1/2$^-$ in the negative parity.  As a matter of
fact, these states are almost degenerate in $^9_\Lambda$Be, because the 
$\alpha\Lambda$ $\ell$$\cdot$s force is very small\cite{lamb}.
The $\ell\cdot$s force makes $\alpha$n 
p$_{3/2}$ and p$_{1/2}$ interactions different.  

Because two-body $\alpha$n p$_{3/2}$ interaction is stronger, 
the three-body state 
with a larger $\alpha$n p$_{3/2}$ component has a lower resonance energy.
The  resonance energy for 5/2$^+$ is lower than 3/2$^+$ as can be 
seen in Table 1.  In the positive parity states, higher spin state
5/2$^+$ in this case, has lower resonance energy than the other.  
Note that the width is considerably small at the same time.
We will see the same phenomena throughout this paper.  Among two levels, which
would be degenerate without an $\alpha$n $\ell$$\cdot$s force, the one with
lower resonance energy always has a smaller width.  
The fifth excited state, 3/2$^+$, is located just above the $\alpha\alpha$ 
$\ell$=2 thresholds with a large width.  
Reducing the $\alpha$n p-wave 
interaction, we find the position of this resonance hardly changes any more, 
although the width becomes larger.  The resonance energy of this state is 
already maximum.  Resonance energies, in general, are constrained by the
related threshold.

The argument of the rotational band predict the existence of 9/2$^+$ and
7/2$^+$\cite{oer}.  We find them with ({\it E}, $\it \Gamma$) = 
(6.65, 2.46) MeV and 
({\it E}, $\it \Gamma$) = (8.13, 6.32) MeV, respectively.  The 
$\ell$$\cdot$s force
give a similar difference both in {\it E} and {$\it \Gamma$} to the pair 
(9/2$^+$,7/2$^+$) as the pair (5/2$^+$, 3/2$^+$).  The 9/2$^+$ was found in NM.  
The newly predicted resonance 7/2$^+$ introduced a confusion into the 
interpretation of the experiment as we will discuss later.   We summarize higher
lying resonances predicted by the present calculation in Table 2 together with
corresponding experiments and Arai's calculation.  

Among the first negative parity pair (3/2$^-$, 1/2$^-$), which would be 
degenerate without $\alpha$n $\ell$$\cdot$s force, the former has lower
energy.   Little influence of $\alpha$n p$_{3/2}$-wave 
interaction is expected in 1/2$^-$, because the other $\alpha$ particle
is in d wave relative to the interacting p$_{3/2}$ pair.  
The situation is different in the second negative parity pair (5/2$^-$, 
7/2$^-$).  We find 5/2$^-$ already in Table 1.  On the contrary, 7/2$^-$ has 
a resonance energy large as can be seen in Table 2 .  Not only the resonance 
energy, but also the width
is considerably larger in 7/2$^-$.   Again
they are correlated in the pair.  Both states 5/2$^-$ as well as 7/2$^-$
are in reasonable agreement with experiments SC and NM.

The width of 5/2$^-$ is extremely
small.  It is less than 10$^{-3}$ in our calculation which is consistent with
the experiment.    
Extremely small width indicates that the effect of the opening channel
is small; {\it i.e.}, it is a quasi bound state.  
The 5/2$^-$ level is a high spin level, which is clear from the 
argument of the degeneracy, and has small influence from opening $\alpha\alpha$ 
$\ell$=0 channel.  Reducing the p-wave $\alpha$n interaction, we trace the
resonance energy and the width in 5/2$^-$.  We find the width is reasonably 
large only after the resonance energy is above the $\alpha\alpha$ $\ell$= 2 
channel threshold.  

Since the 5/2$^-$ resonance is so strong as a quasi bound state, we expect
higher spin state 9/2$^-$ in the same rotational band.  We find it at 
({\it E, $\it \Gamma$}) = (10.36, 1.14) MeV.  We identify it $^9$Be(11.28) 
because of
the energy and the width.No level assignment are given in SC, while 7/2$^+$
has assigned in NM.  The discussion given above in the 7/2$^+$ 
is strongly against to this assignment.  A knowledge of the width is 
very critical for the spin-parity assignment.  

The most serious disagreement found in SC and NM is less evident levels 
around 6.7 MeV as summarized in Table 2.  A 7.94 MeV resonance is reported in SC
with $\it \Gamma \sim$ 1 MeV.  Very few experiments in SC treated this resonance.
NM did not find this resonance or at least did not find $\it \Gamma$ reasonable.
Our calculation predict a resonance at 8.13 MeV with extremely large $\it \Gamma$
(6.32 MeV).  Having so large width, it is found in few experiments.  An
influence of this resonance could be seen even below the bump at 7.6 MeV as a 
tail.  NM actually find something necessary to fill the gap at the lower tail of 
the bump.  They introduced a new resonance at 5.59 MeV which, very probably, 
is a tail of the 7/2$^+$ resonance.  

A comment should be given here.  Although the existence of the second level
(3/2)$_2$ at 5.59 MeV has not been reported experimentally except for NM, it is 
predicted by a 
few theoretical works as is shown by Arai {\it et al}.  However, it should be 
noted 
the position of (3/2)$_2$ reported is very close to the $\alpha \alpha$ $\ell=2$
threshold.  The threshold effect may cause some disturbance in the experiment.  
Although the present calculation suggest that the structure found in NM may
be a tail of higher broad resonance combined with the threshold effect, more
careful studies both in theory and experiment will be necessary with careful
treatment of the threshold.  It would be important to clarify the position
of the energy level as well as the width relative to the singularity caused 
by $\alpha \alpha$ $\ell=2$ resonance.  Even the Riemann sheet in the
calculation may be different from corresponding measurement.  In that case,
we should not regard the theoretical prediction to be confirmed by the
experiment.  

We can not find any signals of resonances at higher energies than the
9/2$^-$.  Therefore, experimentally 
observed higher energy resonances have structures other than 
$\alpha$$\alpha$n.   We do not find it necessary to introduce second levels
in low spin states to explain existing data below 11.28 MeV.  
They will be found above it. 
Because they do not have an $\alpha$$\alpha$n structure, the
decay widths are expected to be small.  Experimentally observed resonances above
11.28 MeV have smaller width.  They support the present interpretation.

In conclusion, we found 10 levels within the framework of the three-body model.
It should be noted that the present paper introduces a new viewpoint 
on three-body resonances in general, which has not been recognized clearly.  
The spin-parity can be assigned with a help of the width.  
So far, works have been done within the framework of the three-body model
without considering complicated threshold effects correctly.  Resonance positions
are strongly influenced by these thresholds, because resonances decay through
these channels.  The threshold effect is definitely given in the present 
framework because it is the input of the three-body equation through the
rigorous solution of the two-body subsystem.  As a price, we started with
phenomenological two-body interactions which is easier to treat.  In that 
framework, we only get well developed cluster levels.  

Further works on this direction will unveil structures of few-body resonances. 
In a subsequent paper\cite{lamb}, we discuss $^9_\Lambda$Be within a 
similar model.  Further works on three-body resonant systems will be
done carefully considering the relations of the resonance poles and 
singularities.  On that process, the knowledge of level structure of $^9$Be 
will become more definite. 

\acknowledgements

The present work was done when the author was at Lisbon.  He is grateful to 
A.C. Fonseca and E. Cravo for discussions and their kind hospitality during his
stay at Lisbon.  He also would like to express his thanks to H. Kamada for 
his interest and discussions.

\begin{center}
    \begin{tabular}{ccccccccccc}
\multicolumn{11}{c}{\bf Table-1}  \\
\multicolumn{11}{c}{Experimentally well established low lying levels }  \\
\hline
\hline
 & & \multicolumn{2}{c}{ Present } & & \multicolumn{2}{c}{ Arai {\it et al.} }
 & & & \multicolumn{2}{c}{ SC and NM }   \\
 & & {\it E} (MeV) & $ { \it \Gamma}$ (MeV)   &  & {\it E} (MeV) & 
 $ { \it \Gamma}$ (MeV)  & &
 & {\it E} (MeV) & $ { \it \Gamma} $ (MeV)  \\
\hline
$3/2^-$ & & 0 & 0  & & 0 & 0  & & &  0 & 0 \\
$1/2^+$ & & \multicolumn{2}{c}{ (virtual state) }  & & & & & & 1.684 & 0.217
 \\
$5/2^-$ & & 2.33 & $< 10^{-3}$ & & 2.27 & 0.001 & & & 2.429 & 0.00077  \\
$1/2^-$ & & 2.82 & 0.50 & & 2.63 & 0.46 & & & 2.780 & 1.08  \\
$5/2^+$ & & 3.91 & 0.35 & & 3.41 & 0.6 & & & 3.049 & 0.282 \\
$3/2^+$ & & 4.95 & 1.95 & & 4.7 & 1.6 & & & 4.704 & 0.743   \\
\hline
\hline
\end{tabular}
\end{center}
\medskip
\begin{center}
    \begin{tabular}{ccccccccccccccccc}
\multicolumn{17}{c}{\bf Table-2}  \\
\multicolumn{17}{c}  { Higher lying resonances }  \\
\hline
\hline
Energy &   \multicolumn{3}{c}{ Present } & & \multicolumn{3}{c}{ Arai {\it et al.} }
  & & \multicolumn{4}{c}{ SC } & & \multicolumn{3}{c}{ NM }  \\

(MeV)  & {\it E}  & $ { \it \Gamma} $ & $J^\pi$ &  & {\it E}  & 
 $ { \it \Gamma} $  & $J^\pi$ & &
 & {\it E}  & $ { \it \Gamma} $ & $J^\pi$ & & {\it E} & 
 $ { \it \Gamma} $ & $J^\pi$ \\
\hline
 $<$ 6.7  &  \multicolumn{2}{c}{tail of the ${7\over 2}^+$} & & & 4.3 & 0.8 & ${3 \over  2}^-$ & 
 & & \multicolumn{3}{c}{None} & & 5.59 & 1.33 & ${3\over 2}^-$
  \\
 $\sim$ 6.7   & 6.68 & 1.98 & ${7\over 2}^-$ &  & 6.46 & 1.2 & ${7\over 2}^-$ & & & 6.76 & 1.54 & ${7\over 2}^-$ & & 6.38 & 1.21 & ${7\over 2}^-$ \\
 $\sim$ 6.7 & 6.65 & 2.46 & ${9\over 2}^+$ & & 6.3 & 2.9 & ${9\over 2}^+$ & & 
 & & None & & & 6.76 & 1.33 &
  ${9\over 2}^+$ \\
$>$ 6.7   &  8.13 & 6.32 & ${7 \over 2}^+ $ & & 7.9 & 2.1 & ${5 \over 2}^-$ & &  & 7.94 & $\sim 1$  & ${1 \over 2}^-$ & & \multicolumn{3}{c}{None}   \\
11.28  & 10.36 & 1.14 & ${9 \over 2}^-$ &  & \multicolumn{3}{c}{None} & & &
11.28 & 0.575 &  & & 11.28 & 1.14& ${7 \over 2}^+$ \\
\hline
\hline
\end{tabular}
\end{center}
\end{document}